\documentclass[10pt,a4paper,twoside]{article}
\usepackage{epsfig}
\usepackage{baltlat5}
\usepackage{wrapfig}
\begin{document}
\ \
\vspace{0.5mm}

\setcounter{page}{1}
\vspace{5mm}

\titlehead{Baltic Astronomy, vol.\ts 14, XXX--XXX, 2005.}

\titleb{SUBDWARF B BINARIES IN THE EDINBURGH-CAPE SURVEY}


\begin{authorl}
\authorb{L. Morales-Rueda}{1}
\authorb{P.F.L. Maxted}{2}
\authorb{T. R. Marsh}{3}
\authorb{D. Kilkenny}{4} and
\authorb{D. O'Donoghue}{4}
\end{authorl}

\begin{addressl}
\addressb{1}{IMAPP, Department of Astrophysics, Radboud University
  Nijmegen, The Netherlands}
\addressb{2}{School of Chemistry and Physics, Keele University, United Kingdom}
\addressb{3}{Department of Physics, Warwick University, United Kingdom}
\addressb{4}{South African Astronomical Observatory, South Africa}
\end{addressl}

\submitb{Received 2005 July 1}

\begin{abstract}
We give an update of the results of a campaign to obtain orbital
solutions of subdwarf B stars from the Edinburgh-Cape survey (Stobie
et al. 1997). To date we have obtained blue spectra of 40 subdwarf B
stars from the Edinburgh-Cape catalogue using the grating spectrograph
at the 1.9\,m Radcliffe telescope at the South African Astronomical
Observatory. We find that 17 out of these 40 are certain binaries with
a few other objects showing radial velocity variations of small
amplitude. The binary fraction found in our sample, after correcting
for our binary detection efficiency, is 48\%.  We have secured the
orbital parameters for 4 of the 17 systems and narrowed down the
orbits of another 7 to a small range of periods.

Out of the four subdwarf B binaries for which we have determined the
orbital solution, three have orbital periods that, according to
population synthesis studies by Han et al. (2003), suggest they have
been formed via a common envelope ejection channel. The masses of the
companions, assuming a canonical mass of 0.5\,M$_{\odot}$ for the
subdwarf B star, suggest that they are probably white dwarfs.  We
observed the shortest period binary (3\,h) of the three, to search for
indications of modulation in the lightcurve due to irradiation of the
companion by the subdwarf B star. No indications of reflection effect
were found confirming that the companion is indeed a white dwarf. The
fourth system with measured orbital parameters shows an orbital period
that could correspond to a subdwarf B binary formed either via the
common envelope ejection channel or the stable Roche Lobe overflow
channel.

The aim of the this study: to obtain an independent, statistically
significant sample of subdwarf B binaries, with solved orbits, based
purely upon the Edinburgh-Cape survey to avoid the uncertain biases of
the Palomar-Green and other surveys, is underway.

\end{abstract}

\vskip4mm

\newpage

\resthead{SdB binaries in the EC Survey}{L.~Morales-Rueda et al.}

\sectionb{1}{THE BINARY FRACTION OF SUBDWARF B STARS}

Maxted et al. (2001) find that 69$\pm$9\%\ of the subdwarf B (sdB)
stars in their observed sample are in binary systems. Napiwotzki et
al. (2004) find a binary fraction of 40\%\ in their SPY (Supernova
type Ia Progenitor Survey) sample.

\vskip1mm

\begin{wrapfigure}[18]{r}[0pt]{62mm}
\vskip-2mm
\vbox{
\vskip2mm
\centerline{\psfig{figure=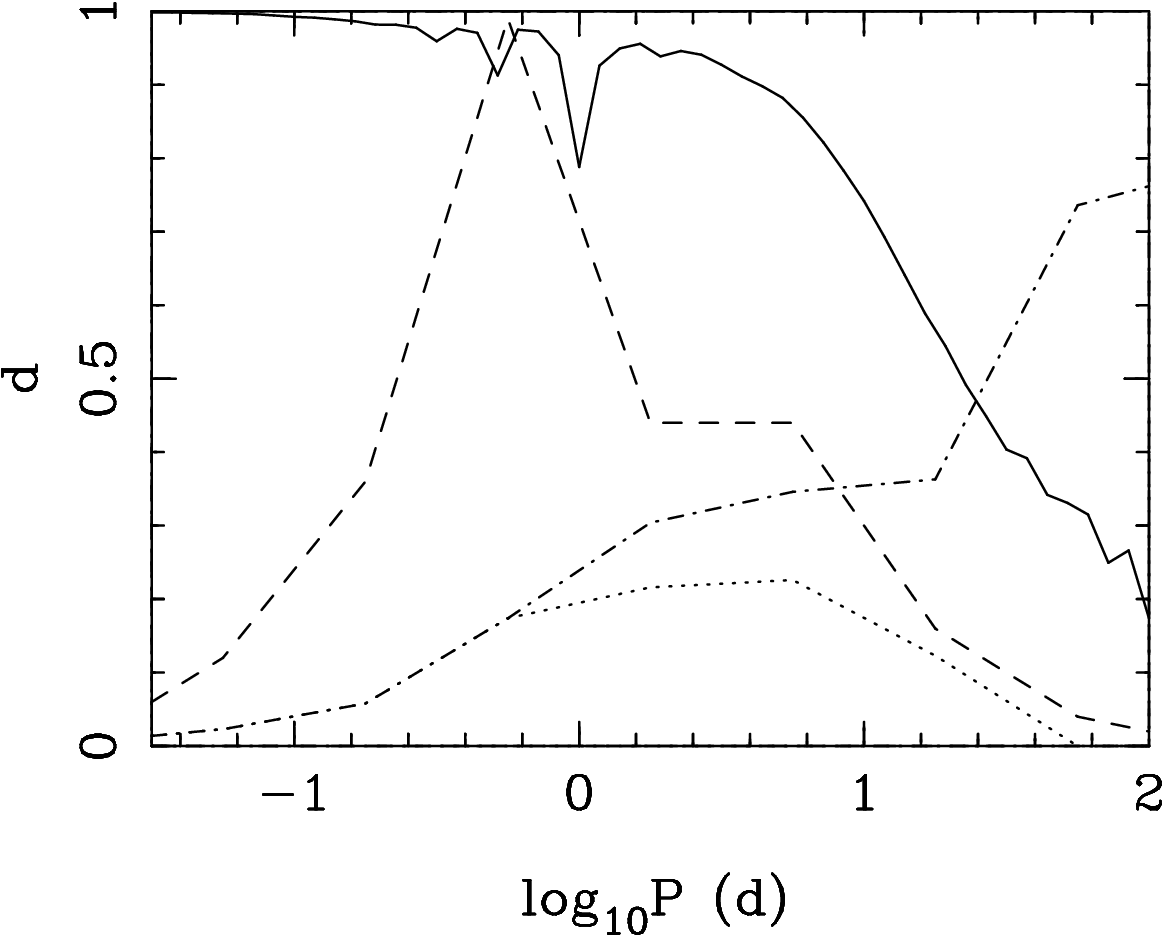,width=50mm,angle=-90,clip=}}
\vskip1mm
\captionc{1}{Detection efficiency as a function of orbital period.}
}
\end{wrapfigure}

Radial velocity measurements of a sample of 40 sdBs from the
Edinburgh-Cape (EC) survey yield 17 certain spectroscopic
binaries. Radial velocities were measured by fitting a model line
profile to H$\beta$ and H$\gamma$ simultaneously (Morales-Rueda et
al. 2004). To determine the true binary fraction in our sample we need
to compute our detection efficiency, i.e. the probability of detecting
(or not detecting) a binary at a certain orbital period due to the
sampling of the data and the accuracy of the radial velocity
measurements. These probabilities were calculated in a similar way to
those by Maxted et al. (2001) and are shown in Fig.~1 (solid
line). For comparison we are also plotting in Fig.~1 the observed
orbital period distribution (dashed line) and the theoretical
distribution, considering both the common envelope (CE) ejection and
the Roche lobe overflow (RLOF) channels (dash-dotted line) and only
the CE ejection channel (dotted line). These distributions will be
discussed again in Section~3. We find that for orbital periods up to 1
day our average detection efficiency is 87\%\ which gives a binary
fraction for our sample of 49$\pm$8\%. The observed distribution peaks
at $\log_{10}P = -0.1$ and the theoretical distribution (only
considering binaries formed through the CE ejection channel) peaks at
$\log_{10}P = 0.6$ where the detection efficiency is 90\%\ in which
case the binary fraction of our sample is 47$\pm$8\%. We find that
this number agrees better with the binary fraction found by Napiwotzki
et al. (2004) than with that found by Maxted et al. (2001).

\vskip1mm

Napiwotzki et al. (2004) suggested that this discrepancy in binary
fraction could be due to the fact that the SPY sdB sample belongs
mainly to the thick disk and the halo, whereas the PG sample studied
by Maxted et al. (2001) comes from the thin disk. In the case of our EC
sample, we expect most of the sdBs to belong to the thin disk
population which indicates that the reason for the discrepancy in
binary fraction is due to something else, probably to low number
statistics.

It is also worth noticing how our detection probability decreases with
longer period systems (less than 50\%\ above 25 days). A longer time
baseline is one of the requirements to increase our sensitivity in
this period range.

\sectionb{2}{ORBITAL SOLUTIONS}

We find the orbital solutions for four of the systems observed,
EC00404$-$4429, EC02200$-$2338, EC12327$-$1338, and
EC12408$-$1427. The orbital solutions for EC00404$-$4429 and
EC02200$-$2338 were already presented by Morales-Rueda et al. (2005)
and are given in parenthesis in the following paragraphs. The orbital
solutions for EC12327$-$1338, and EC12408$-$1427 are given in
Table.~1.

{\bf EC00404$-$4429} (P = 0.12834(4)\,d, M$_2$min = 0.32 M$_{\odot}$):
Its orbital period places it in the group of sdB binaries formed via
the CE ejection channel (see right panel of Fig.~2). The minimum mass
of the companion, assuming the canonical mass of 0.5\,M$_{\odot}$ for
the sdB, indicates that the companion is probably a white dwarf.  We
have looked for indications of a reflection effect on the companion of
this system as it is the shortest period binary of our sample and
found no significant reflection effect. This confirms that the
companion is a white dwarf. The system must have formed therefore via
the second CE ejection channel (Han et al. 2003)

{\bf EC02200$-$2338} (P = 0.8022(7)\,d, M$_2$min = 0.39 M$_{\odot}$)
\& {\bf EC12327$-$1338}: Their orbital periods place them in the group
of sdB binaries formed via the CE ejection channel. The minimum masses
of the companions, assuming the canonical mass for the sdB star,
indicate that the companions are probably white dwarfs.

{\bf EC12408$-$1427}: The orbital period of this sdB binary is
consistent with the binary having been formed either via the CE
ejection channel or via the RLOF channel. The minimum mass of the
companion is compatible with both a white dwarf or a main sequence star.

\begin{center}
\vbox{\norm
\tabcolsep=15pt
\begin{tabular}{lcc}
\multicolumn{3}{c}{\parbox{90mm}{
{\bf \ \ Table 1.}{\ Orbital solution for two sdB binaries.
  $\gamma$ is the systemic velocity, K is the radial velocity
  semiamplitude, and the 1 and 10 per cent rows give the probability that
  the true period lies further than 1 and 10 per cent (respectively)
  from the given value. The numbers given are the $\log_{10}$ of the
  probabilities.}}}\\
\tablerule
\multicolumn{1}{c}{} & \multicolumn{1}{c}{EC12327-1338} &
\multicolumn{1}{c}{EC12408-1427} \\
\tablerule
Period (d) & 0.363221(1) & 0.90243(1) \\
HJD$_0$ (d) & 2452728.153(1) & 2452732.068(5)\\
$\gamma$ (km s$^{-1}$) & $-$6.44 $\pm$ 1.74 & $-$52.02 $\pm$ 1.19\\
K (km s$^{-1}$) & 124.30 $\pm$ 2.55 & 58.90 $\pm$ 1.55\\
M$_2$min (M$_{\odot}$) & 0.38 & 0.21\\
$\chi^{2}_{\rm reduced}$ & 1.9 & 0.8\\
2nd best alias (d) & 0.369281(1) & 9.493(1)\\
$\Delta\chi^2$  & 33 & 38\\
n & 15 & 29\\
1 per cent & $-$7.34 & $-$6.89\\
10 per cent & $-$11.58 & $-$6.96\\
Systematic error (km s$^{-1}$) & 2 & 2\\
\tablerule
\end{tabular}
}
\end{center}

\sectionb{3}{ORBITAL PERIOD DISTRIBUTION}

Theory predicts that most sdB stars should be in long period binaries
(Han et al. 2003). They would have formed via a stable Roche Lobe
overflow channel (dashed line in right panel of Fig.~2) and have main
sequence companions. This long period population is missing from the
observations shown in the left panel of Fig. 2. At present only two
long period sdBs candidates are known. This is probably caused by
biases in the observed sample: 1. early type companions will swamp the
light of the sdB star, 2. long period binaries will show smaller
amplitude radial velocities thus higher resolution spectra is needed
to find them, 3. longer time baselines are required to measure periods
of a few hundred days.

Biases numbers 2 and 3 affect directly the binary detection efficiency
curves presented in Section~1. This explains the differences between
the observed and the predicted orbital period distributions at long
orbital periods.

\vbox{
\begin{minipage}{5cm}
\vskip5mm
\hskip-5mm
{\psfig{figure=sdBbinaries_hist_Jul05_bw.ps,width=50mm,angle=-90,clip=}}
\end{minipage}
\begin{minipage}{5cm}
{\psfig{figure=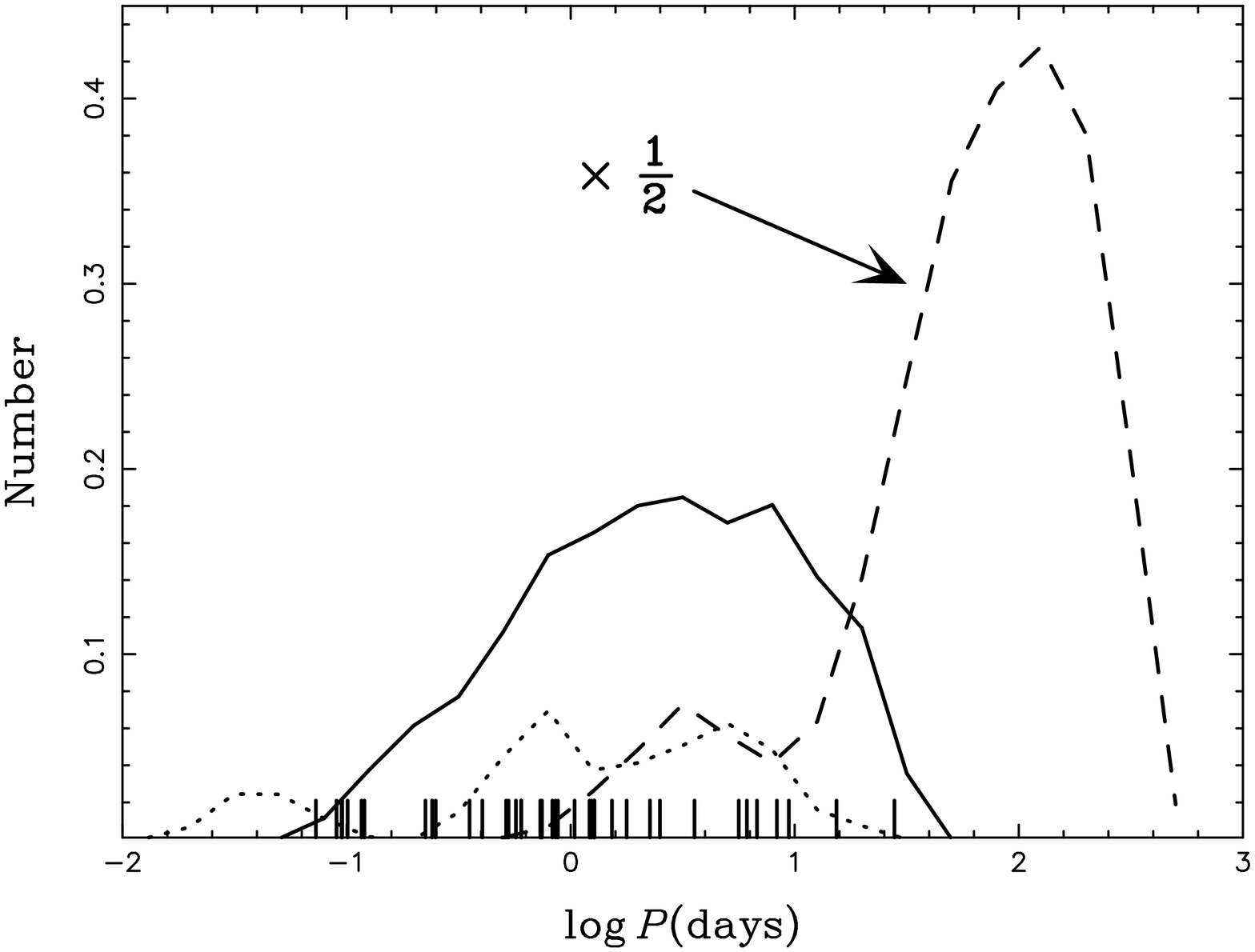,width=55mm,angle=-90,clip=}}
\end{minipage}

\vskip1mm \captionc{2}{Left panel: Observed orbital period
distribution of sdB binaries. Light grey: unknown companion type, dark
grey: main sequence companions, black: white dwarf companions. Right
panel: theoretical orbital period distribution taken from Han et
al. (2003). Dotted line: CE channel sdBs with white dwarf companions,
solid line: CE channel sdBs with main sequence companions, dashed
line: stable Roche Lobe overflow channel sdBs with main sequence
companions.}  }

\vskip 2mm

With 40 EC sdBs observed and 17 binaries found, this study is well on
its way. More observations are required to obtain the statistically
significant sample, based only on EC sdBs, that we seek.

\vskip2mm 

ACKNOWLEDGEMENTS. This paper uses observations made at the South
African Astronomical Observatory (SAAO). LMR is supported by NWO-VIDI
grant 639.042.201 to P.J. Groot. The authors would like to thank the
Leids Kerkhoven-Bosscha Fonds for providing funding to attend this
meeting.

\goodbreak

\References

\refb
Han~Z., Podsiadlowski~Ph., Maxted~P.~F.~L., Marsh~T.~R., 2003, MNRAS,
341, 669

\refb
Maxted~P.~F.~L, Heber~U., Marsh~T.~R., North~R.~C., 2001, MNRAS, 326,
  1391

\refb Morales-Rueda~L., Maxted~P.~F.~L., Marsh~T.~R., 2004,
Ap\&SS, 291, 299

\refb Morales-Rueda~L., Maxted~P.~F.~L., Marsh~T.~R. et al. 2005, 
D.~Koester \& S.~Moehler Eds, ASP Conf. Ser. 334, 333

\refb
Napiwotzki~R., et al., 2004, Ap\&SS, 291, 321

\refb
Stobie~R.~S., et al., 1997, MNRAS, 287, 848

\vskip1mm

\end{document}